\documentclass[11pt,a4paper]{article}
\usepackage{bbm}
\pdfoutput=1
\usepackage{jheppub}
\usepackage{amsmath}
\usepackage{epsfig}
\usepackage{amssymb}
\usepackage{graphics}
\usepackage[active]{srcltx}
\usepackage{epstopdf}
\usepackage{subfigure}

\newcommand{\be}{\begin{equation}}
\newcommand{\ee}{\end{equation}}
\newcommand{\bea}{\begin{eqnarray}}
\newcommand{\eea}{\end{eqnarray}}
\newcommand{\bean}{\begin{eqnarray*}}
\newcommand{\eean}{\end{eqnarray*}}
\newcommand{\nn}{\nonumber \\}

\def\W #1{\widetilde{#1}}

\def\eref#1{(\ref{#1})}

\def\d{{\rm d}}

\def\a{{\alpha}}

\def\b{{\beta}}

\def\d{\partial}

\def\eps{\epsilon}

\def\Label#1{\label{#1}%
  \smash{\hbox to0pt{\raise1ex\hbox{\tiny[#1]}\hss}}}

\title{Note on differential operators, CHY integrands, and unifying relations for amplitudes}
\author[a]{Kang Zhou,}
\author[b,c]{Bo Feng\footnote{The unusual ordering of authors instead of the
standard alphabetical one is for postdocs and young researchers to get proper recognition of contributions under
the current out-dated practice in China.}}

\affiliation[a]{Center for Gravitation and Cosmology, College of Physical Science and Technology, Yangzhou University,\\
 No.180, Siwangting Road, Yangzhou, 225009, P.R. China.}
\affiliation[b]{Zhejiang Institute of Modern Physics, Department of Physics,
  Zhejiang University,\\ No.38, Zheda Road, Hangzhou, 310027, P.R. China.}
\affiliation[c]{Center of Mathematical Science,
  Zhejiang University,\\ No.38, Zheda Road, Hangzhou, 310027, P.R. China.}

\emailAdd{zhoukang@yzu.edu.cn}
\emailAdd{fengbo@zju.edu.cn}

\date{\today}
\abstract{An elegant unified web for amplitudes of various theories was given by Cachazo, He and Yuan in the CHY framework a few years ago. Recently, similar web has also been constructed by Cheung, Shen and Wen, which
relies on a set of differential operators. In this note, by acting these differential operators on CHY-integrands systematically, we have established the relation between these two approaches.
Thus, amplitudes for all theories which have CHY representations, include  gravity theory, Einstein-Yang-Mills theory, Einstein-Maxwell theory, pure Yang-Mills theory, Yang-Mills-scalar theory, Born-Infeld theory,
Dirac-Born-Infeld theory and its extension, bi-adjoint scalar theory, $\phi^4$ theory, non-linear sigma model,
as well as special Galileon theory, have been included in the unified web rooted from  gravity theory.
}


\begin{document}

\maketitle \flushbottom

\section{Introduction}
\label{secintro}

The unification of different theories is always one of interesting problems in theoretical physics.
The modern researches on S-matrix have exhibited amazing structures within amplitudes of gauge and gravity theories,
such as the Kawai-Lewellen-Tye (KLT) relations \cite{KLT}, Bern-Carrasco-Johansson (BCJ) color-kinematics duality \cite{Bern:2008qj,Bern:2010ue,Bern:2010yg}, which
are invisible in the traditional Lagrangian formulism  of quantum field theory. These discoveries hint the existence
of some long hidden unifying relations for on-shell amplitudes. A strong evidence for the marvelous unity among amplitudes of different theories  has been spelled out in
\cite{Cachazo:2014xea} by using the CHY formulae \cite{Cachazo:2013gna,Cachazo:2013hca,Cachazo:2013iea,Cachazo:2014nsa,Cachazo:2014xea}. More explicitly, different
theories are defined by different CHY-integrands, while they found that CHY-integrands for a wide range of theories can be generated from the CHY-integrand for gravity theory\footnote{Here the
 gravity theory has to be understood in a generalized version,
i.e., Einstein gravity theory couples to a dilaton and two-forms.}, through the so called compactifying, squeezing
as well as the generalized dimensional reduction procedures \cite{Cachazo:2014xea}.

Recently, Cheung, Shen and Wen discovered similar unifying relations for
on-shell tree-level amplitudes of a variety of theories from a different angle:
by acting
some  Lorentz and gauge invariant differential operators, one can
transmute the physical amplitude of a theory into the one of another theory
\cite{Cheung:2017ems}. In their unified web, amplitudes of various
theories include Einstein-Yang-Mills theory, Einstein-Maxwell theory, Born-Infeld theory,
Dirac-Born-Infeld theory, special Galileon theory,  non-linear sigma
model, as well as bi-adjoint scalar theory, can be generated by
transmuting the amplitudes of gravity theory. The role
of these differential operators has been understood and checked from various
angels, such as several explicit examples, factorization property, double copy structure, soft
behavior, etc.

Since the similar unified webs  for amplitudes of various theories have
been given both in  \cite{Cachazo:2014xea} and \cite{Cheung:2017ems},
it is very natural to investigate the relation among these two
different approaches. In this note, we will establish the exact
relation through the CHY formulae
\cite{Cachazo:2013gna,Cachazo:2013hca,Cachazo:2013iea,Cachazo:2014nsa,Cachazo:2014xea}.
Tree-level amplitudes in the CHY formulae are represented as
integrals over auxiliary variables as
\bea
{\cal A}_n=\int d\mu_n\,{\cal I}^{\rm CHY}\,,~~~~\label{CHY-0}
\eea
where the auxiliary variables are localized by constraints from the so-called scattering equations which depend on the
external momenta. In this formulae, the measure part $d\mu_n$ is universal for all theories, while different theories are defined by the so called CHY-integrands ${\cal I}^{\rm CHY}$.
Based on this fact, the basic idea of the note can be described as following. Since differential operators
discussed in this note are defined through Lorentz invariants include polarization vectors of external particles
such as $\eps_i\cdot \eps_j$ and $\eps_i\cdot k_j$, they are commutable\footnote{We want to remark that
in \cite{Cheung:2017ems}, differential operators such as $\d_{k_i\cdot k_j}$ have also been
discussed. However, these operators will interact with scattering equations, thus we will not use them
in this note.} with the integral $\int d\mu_n$
over auxiliary variables.
Therefore, converting an amplitude is equivalent to converting the CHY-integrand. More explicitly, if two amplitudes ${\cal A}_\a$ and ${\cal A}_\b$ are related by an operator ${\cal O}$ as ${\cal A}_\a={\cal O}{\cal A}_\b$, analogous relation ${\cal I}^{\rm CHY}_\a={\cal O}\,{\cal I}^{\rm CHY}_\b$ for integrands
must hold, and vice versa. Thus, one can derive the unifying relations systematically
by acting operators on CHY-integrands.

Applying differential operators on CHY-integrands, we will re-derive all unifying relation in \cite{Cheung:2017ems}. We will also define new
operators which are composed of basic trace operators, to generate amplitudes of theories having not been mentioned in \cite{Cheung:2017ems}.
Then all amplitudes which have CHY representations in \cite{Cachazo:2014xea} can be bringed into the picture of unification: they are generated from the amplitudes
of gravity theory via several operators. Other relations among amplitudes indicated by these operators will also be discussed.

The remainder of this paper is organized as follows. In \S\ref{secreview}, we give a brief introduction of the Pfaffian and the CHY formulae which are crucial for subsequent discussions. In \S\ref{secbasic}, we study the effects of  three basic operators when acting them on the building blocks of CHY-integrands.
 Then, in \S\ref{secpro} we will consider the effects of operators built by these basic operators. The unified web and other relations for amplitudes will be presented in \S\ref{secunify}. Finally, we end with a summary and discussions in \S\ref{secconclu}.

\section{Review of Pfaffian and CHY formulae}
\label{secreview}

For reader's convenience, we will briefly discuss the definition of Pfaffian, and rapidly review the CHY formulae.

\subsection{Definition of Pfaffian}

The definition of Pfaffian is essential for the work in this note.
For a $2n\times 2n$ skew symmetric matrix $S$, Pfaffian is defined as
\bea
{\bf Pf}S={1\over 2^n n!}\sum_{\sigma\in S_{2n}} {\bf sgn}(\sigma)\prod_{i=1}^n\,a_{\sigma(2i-1),\sigma(2i)}\,,~~~\label{pfa-1}
\eea
where $S_{2n}$ is the permutation group of $2n$ elements and ${\bf sgn}(\sigma)$ is the signature of $\sigma$.
More explicitly, let $\Pi$ be the set of all partitions of $\{1,2,\cdots, 2n\}$ into pairs without regard to the order.
An element $\a$ in $\Pi$ can be written as
\bea
\a=\{(i_1,j_1),(i_2,j_2),\cdots,(i_n,j_n) \}\,,
\eea
with $i_k<j_k$ and $i_1<i_2<\cdots<i_n$. Now let
\bea
\pi_{\a} = \left(
         \begin{array}{c}
           ~~~1~~~ 2~~~3~~~4~~\cdots~2n-1~~2n~~ \\
           \,\,i_1~~j_1~~i_2~~j_2~~\cdots~~~i_n~~~~~~j_n \\
         \end{array}
       \right)
\eea
be the corresponding permutation of the partition $\a$. If we define
\bea
S_{\a}={\bf sgn}(\pi_{\a})\,a_{i_1j_1}a_{i_2j_2}\cdots a_{i_nj_n}\,,
\eea
then the Pfaffian of the matrix $A$ is given as
\bea
{\bf Pf}S=\sum_{\a\in\Pi}S_{\a}\,.~~~~~\label{pfa}
\eea
Both representations \eref{pfa-1} and \eref{pfa} will be used later.
From the \eref{pfa} one can observe that in every term $S_{\a}$
of the Pfaffian, each number of $\{1,2,\cdots,2n\}$, as the subscript of the matrix element, will appear once and only once. This observation is simple but useful for latter discussions.

\subsection{CHY formulae}

With the definition of Pfaffian described above, now we can introduce the CHY formulae \cite{Cachazo:2013gna,Cachazo:2013hca,Cachazo:2013iea,Cachazo:2014nsa,Cachazo:2014xea}.
In the CHY formulae, tree level amplitudes for $n$ massless particles arise from a multi-dimensional contour integral over
the moduli space of genus zero Riemann surfaces with $n$ punctures, ${\cal M}_{0,n}$. It can be expressed as
\bea
{\cal A}_n=\int d\mu_n\,{\cal I}_L(\{k,\epsilon,z\}){\cal I}_R(\{k,\W\epsilon,z\})\,,~~~~\label{CHY}
\eea
which possesses the M\"obius ${\rm SL}(2,\mathbb{C})$ invariance. Here $k_i$, $\epsilon_i$ and $z_i$ are the momentum, polarization vector, and puncture location for $i^{\rm th}$
particle, respectively. The measure is defined as
\bea
d\mu_n\equiv{d^n z\over{\rm vol}\,{\rm SL}(2,\mathbb{C})}\prod_i{'}\delta({\cal E}_i)\,.
\eea
The $\delta$-functions impose the scattering equations
\bea
{\cal E}_i\equiv\sum_{j\in\{1,2,\ldots,n\}\setminus\{i\}}{s_{ij}\over z_{ij}}=0\,,
\eea
where $s_{ij}\equiv(k_i+k_j)^2$ is the Mandelstam variable,
and $z_{ij}\equiv z_i-z_j$. The scattering equations define the map from the space of
kinematic variables to ${\cal M}_{0,n}$, and fully localize the integral on
their solutions.

The integrand in \eref{CHY} depends on the theory under consideration, and
carries all kinematical information of external particles. For any theory known to have a CHY representation, the corresponding integrand can be split into two
parts ${\cal I}_L$ and ${\cal I}_R$, as can be seen in \eref{CHY}. Either of them are weight-$2$ for each variable $z_i$
under the M\"obius transformation. We list integrands for various
theories as in Table \ref{tab:theories}\cite{Cachazo:2014xea}\footnote{For theories contain gauge or flavor groups, we only show
the integrands for color-ordered partial amplitudes instead of full ones.}.
\begin{table}[!h]
    \begin{center}
        \begin{tabular}{c|c|c}
            Theory& ${\cal I}_L(k,\epsilon,z)$ & ${\cal I}_R(k,\W\epsilon,z)$ \\
            \hline
            gravity theory & ${\bf Pf}'\Psi$ & ${\bf Pf}'{\Psi}$ \\
            Einstein-Yang-Mills & ${\cal C}_{{\rm Tr}_1}\cdots{\cal C}_{{\rm Tr}_m}{\sum_{\{i,j\}}}'{\cal P}_{\{i,j\}}(n,l,m)$ & ${\bf Pf}'{\Psi}$ \\
            pure Yang-Mills & ${\cal C}_n(\sigma)$ & ${\bf Pf}' \Psi$ \\
            Einstein-Maxwell & ${\bf Pf}'[\Psi]_{n-2m,2m:n-2m}{\bf Pf}[X]_{2m}$ & ${\bf Pf}'{\Psi}$ \\
            Einstein-Maxwell(photon with flavor) & ${\bf Pf}'[\Psi]_{n-2m,2m;n-2m}{\bf Pf}[{\cal X}]_{2m}$
            & ${\bf Pf}'{\Psi}$ \\
            Born-Infeld & $({\bf Pf}'A)^2$ & ${\bf Pf}' \Psi$ \\
            Yang-Mills-scalar & ${\cal C}_{{\rm Tr}_1}\cdots{\cal C}_{{\rm Tr}_m}{\sum_{\{i,j\}}}'{\cal P}_{\{i,j\}}(n,l,m)$ & ${\cal C}_n(\sigma)$ \\
            Yang-Mills-scalar(special) & ${\bf Pf}'[\Psi]_{n-2m,2m;n-2m}{\bf Pf}[{\cal X}]_{2m}$
            & ${\cal C}_n(\sigma)$ \\
            pure bi-adjoint scalar & ${\cal C}_n(\W\sigma)$ & ${\cal C}_n(\sigma)$ \\
            non-linear sigma model & $({\bf Pf}' A)^2$ & ${\cal C}_n(\sigma)$ \\
            $\phi^4$ & ${\bf Pf}' A\,{\bf Pf} [X]_n$ & ${\cal C}_n(\sigma)$ \\
            extended Dirac-Born-Infeld & ${\cal C}_{{\rm Tr}_1}\cdots{\cal C}_{{\rm Tr}_m}{\sum_{\{i,j\}}}'{\cal P}_{\{i,j\}}(n,l,m)$
            & $({\bf Pf}' A)^2$ \\
            Dirac-Born-Infeld  & ${\bf Pf}'[\Psi]_{n-2m,2m;n-2m}{\bf Pf}[{\cal X}]_{2m}$ & $({\bf Pf}' A)^2$ \\
            special Galileon & $({\bf Pf}'A)^2$ & $({\bf Pf}' A)^2$ \\
        \end{tabular}
    \end{center}
    \caption{\label{tab:theories}Form of the integrands for various theories}
\end{table}

We now explain each ingredient appearing in this table in turn. The $n\times n$ matrixes are defined through
\bea
& &A_{ij} = \begin{cases} \displaystyle {k_{i}\cdot k_j\over z_{ij}} & i\neq j\,,\\
\displaystyle  ~~~ 0 & i=j\,,\end{cases} \qquad\qquad\qquad\qquad B_{ij} = \begin{cases} \displaystyle {\epsilon_i\cdot\epsilon_j\over z_{ij}} & i\neq j\,,\\
\displaystyle ~~~ 0 & i=j\,,\end{cases} \nn
& &C_{ij} = \begin{cases} \displaystyle {k_i \cdot \epsilon_j\over z_{ij}} &\quad i\neq j\,,\\
\displaystyle -\sum_{l=1,\,l\neq j}^n\hspace{-.5em}{k_l \cdot \epsilon_j\over z_{lj}} &\quad i=j\,,\end{cases}
\label{ABCmatrix}
\eea
and
\bea
X_{ij}=\begin{cases} \displaystyle \frac{1}{z_{ij}} & i\neq j\,,\\
\displaystyle ~ ~ 0 & i=j\,,\end{cases} \qquad\qquad\qquad\qquad
{\cal X}_{ij}=\begin{cases} \displaystyle \frac{\delta^{I_i,I_j}}{z_{ij}} & i\neq j\,,\\
\displaystyle ~ ~ 0 & i=j\,.\end{cases}
\eea
where $\delta^{I_i,I_j}$ forbids the interaction between particles with different flavors. To clarify the dimension, we denote the $n\times n$ matrixes $X$ and ${\cal X}$ as $[X]_n$, $[{\cal X}]_n$. The $2n\times2n$ antisymmetric matrix $\Psi$ is given by
\bea\label{Psi}
\Psi = \left(
         \begin{array}{c|c}
           ~~A~~ &  ~~C~~ \\
           \hline
           -C^{\rm T} & B \\
         \end{array}
       \right)\,.
\eea
The reduced Pfaffian of $\Psi$ is defined as ${\bf Pf}'\Psi={(-)^{i+j}\over z_{ij}}{\bf Pf}\Psi^{[i,j]}$,
where the notation $\Psi^{[i,j]}$ means the rows and columns $i$, $j$ of the matrix $\Psi$
have been deleted (with $1\leq i,j\leq n$). It can be proved that this definition is independent of the choice of $i$ and $j$.
Analogous notation holds for ${\bf Pf}'A$.

The definition of $\Psi$ can be generalized to the $(2a+b)\times(2a+b)$
case $[\Psi]_{a,b:a}$  as
\bea
[\Psi]_{a,b:a}=\left(
         \begin{array}{c|c}
           ~~A_{(a+b)\times (a+b)}~~ &  C_{(a+b)\times a} \\
           \hline
            -C^{\rm T}_{a\times (a+b)} & B_{a\times a} \\
         \end{array}
       \right)\,,~~~~\label{psi-aba}
\eea
here $A$ is a $(a+b)\times (a+b)$ matrix, $C$ is a $(a+b)\times a$ matrix, and $B$ is a $a\times a$ matrix.
The definitions of elements of $A$, $B$ and $C$ are the same as before. The reduced Pfaffian ${\bf Pf}'[\Psi]_{a,b:a}$
is defined in the same manner. With the definition of the reduced Pfaffian, one can observe that: {\sl each polarization vector $\epsilon_i$
appears once and only once in each term of the reduced Pfaffian.}

Furthermore, starting from the $2n\times 2n$ matrix $\Psi$, the polynomial ${\cal P}_{\{i,j\}}(n,l,m)$ is defined by
\bea
{\cal P}_{\{i,j\}}(n,l,m)&=&
{\bf sgn}(\{i,j\})\,
z_{i_1 j_1}\cdots z_{i_m j_m}\,
{\bf Pf}'[\Psi]_{n-l,i_1,j_1,\ldots,i_m,j_m: n-l}\nn
&=&-{\bf sgn}(\{i,j\}')\,
z_{i_1 j_1}\cdots z_{i_{m-1} j_{m-1}}\,
{\bf Pf}[\Psi]_{n-l,i_1,j_1,\ldots,i_{m-1},j_{m-1}: n-l}\,,~~~~\label{multi-trace-poly}
\eea
where $i_k<j_k\in{\rm Tr}_k$ and ${\rm Tr}_k$'s are $m$ sets satisfy\footnote{Each set has at least two elements,
so in general we have $l\geq 2m$.}
\bea
{\rm Tr}_1\cup{\rm Tr}_2\cup\cdots\cup{\rm Tr}_m=\{n-l+1,n-l+2,\cdots,n\}\,.
\eea
In the notation $[\Psi]_{n-l,i_1,j_1,\ldots,i_m,j_m: n-l}$, we explicitly write $\{i_1,j_1,\ldots,i_m,j_m\}$
instead of $2m$ to emphasize the locations of $2m$ rows and $2m$ columns in the original matrix $\Psi$.
Two signatures ${\bf sgn}(\{i,j\})$ and ${\bf sgn}(\{i,j\}')$
correspond to partitions $\{(i_1,j_1),\cdots,(i_m,j_m)\}$ and $\{(i_1,j_1),\cdots,(i_{m-1},j_{m-1})\}$
respectively, and one can verify ${\bf sgn}(\{i,j\})={\bf sgn}(\{i,j\}')$.
In the second line of \eref{multi-trace-poly},
the reduced Pfaffian is calculated by removing
rows and columns $i_m$ and $j_m$, and $(-)^{(n-l+2m-1)+(n-l+2m)}=(-)$ have been used.
Under the definition of ${\cal P}_{\{i,j\}}(n,l,m)$ in the second line,
the summation ${\sum_{\{i,j\}}}'{\cal P}_{\{i,j\}}(n,l,m)$ means
\bea
{\sum_{\{i,j\}}}'{\cal P}_{\{i,j\}}(n,l,m)\equiv\sum_{\substack{i_1<j_1\in{\rm Tr}_1\\\cdots\\i_{m-1}<j_{m-1}\in{\rm Tr}_{m-1}}}{\cal P}_{\{i,j\}}(n,l,m)\equiv {\bf Pf}'\Pi \,.~~~\label{Pfa-Pi}
\eea
where the sum is over all possible choices of pairs in each trace subset.
Notice that one can choose to delete rows and columns belong to any ${\rm Tr}_k$ when computing the reduced Pfaffian, and
${\sum_{\{i,j\}}}'{\cal P}_{\{i,j\}}(n,l,m)$ is independent of the choice since it is equal to the reduced Pfaffian of $\Pi$, which is constructed using the squeezing procedure \cite{Cachazo:2014xea}.

Finally, the Parke-Taylor factor for ordering $\sigma$ is given as
\bea
{\cal C}_n(\sigma)={1\over z_{\sigma_1\sigma_2}z_{\sigma_2\sigma_3}\cdots z_{\sigma_{n-1}\sigma_n}z_{\sigma_n\sigma_1}}\,,
\eea
it implies the color order $\{\sigma_1\sigma_2\cdots \sigma_{n-1}\sigma_n\}$ for the partial amplitude.

\section{Basic operators}
\label{secbasic}

In this section, we will consider the effects
of acting three basic differential operators given in \cite{Cheung:2017ems} on the elementary building-blocks of CHY-integrands such as
${\bf Pf}'\Psi$, ${\bf Pf}'[\Psi]_{a,b:a}$, as well as $\sum_{\{i,j\}}'{\cal P}_{\{i,j\}}(n,l,m)$.

\subsection{Trace operator}

The trace operator ${\cal T}_{ij}$ is defined as \cite{Cheung:2017ems}
\bea
{\cal T}_{ij}\equiv\partial_{\epsilon_i\epsilon_j}\,.
\eea
Here $\eps_i\eps_j$ means $\eps_i\cdot \eps_j$ and the differential operator is to take derivative regarding to the
combination $\eps_i\cdot \eps_j$. Similar understanding holds for all operators in this note.
If one apply ${\cal T}_{ij}$ on the reduced Pfaffian  ${\bf Pf}'\Psi$, only terms containing factor $\epsilon_i\epsilon_j$ (i.e.,element $\Psi_{i+n,j+n}$)
provide non-vanishing contributions. Thus
performing the operator ${\cal T}_{ij}$
is equivalent to the replacement
\bea
\epsilon_i\epsilon_j\to1\,,~~~~\epsilon_iV\to 0\,,~~~~\epsilon_jV\to0\,,
\eea
where $V$ denotes vectors $k_l$'s or $\epsilon_{l\neq i,j}$'s, since
$\epsilon_i$ and $\epsilon_j$ appear once and only once in each term
of the reduced Pfaffian respectively. As noted in
\cite{Cheung:2017ems}, the effect is nothing but the dimensional
reduction (or the "compactifying" procedure in
\cite{Cachazo:2014xea}). Thus we arrive at a new matrix $\W\Psi$
satisfies
\bea
{\cal T}_{ij}\,{\bf Pf}'\Psi={\bf Pf}'\W\Psi\,.
\eea
Without lose of generality, one can assume $\{i,j\}=\{n-1,n\}$\footnote{This assumption can be realized by moving lows
and columns. Since $(n+i)^{\rm th}$ row and column will be moved simultaneously while moving $i^{\rm th}$ ones,
the possible $-$ sign will not arise.}, then the new matrix $\W\Psi$ is given by
\bea
\W\Psi = \left(
         \begin{array}{c|c|c}
           ~~A_{n\times n}~~ &  C_{n\times(n-2)} & 0 \\
           \hline
           -C^{\rm T}_{(n-2)\times n} & B_{(n-2)\times(n-2)} & 0\\
           \hline
           0 & 0 & X_{2\times2}\\
         \end{array}
       \right)=\left(
         \begin{array}{c|c}
          [\Psi]_{n-2,2:n-2}& 0\\
           \hline
            0 & [X]_2\\
         \end{array}
       \right)\,.~~~~~\label{result-M}
\eea
The reduced Pfaffian of the matrix $\W\Psi$ can be calculated straightforwardly as
\bea
{\bf Pf}'\W\Psi={\bf Pf}'[\Psi]_{n-2,2;n-2}{\bf Pf}[X]_2\,.~~~~\label{result1}
\eea
Thus, we find
\bea
{\cal T}_{ij}\,{\bf Pf}'\Psi&=&{\bf Pf}'[\Psi]_{n-2,2;n-2}{\bf Pf}[X]_2\,.~~~~\label{result-trace-1}
\eea
Same analysis gives  the result of trace operator acting on generalized matrix $[\Psi]_{a,b:a}$
\bea
{\cal T}_{ij}{\bf Pf}'[\Psi]_{a,b:a}={\bf Pf}'[\Psi]_{a-2,b+2:a-2}{\bf P
f}[X]_2\,.
\eea
Repeating the manipulations,
multiple action of trace operators give following generalization of \eref{result-trace-1} as
\bea
{\cal T}_{i_1j_1}{\cal T}_{i_2j_2}\,{\bf Pf}'\Psi&=&{\bf Pf}'[\Psi]_{n-4,4;n-4}{\bf Pf}[X_1]_2{\bf Pf}[X_2]_2\,,\nn
& &\cdots\nn
{\cal T}_{i_1j_1}{\cal T}_{i_2j_2}\cdots{\cal T}_{i_mj_m}\,{\bf Pf}'\Psi&=&{\bf Pf}'[\Psi]_{n-2m,2m;n-2m}{\bf Pf}[X_1]_2{\bf Pf}[X_2]_2
\cdots{\bf Pf}[X_m]_2\nn
&=&{(-)^m\over
(z_{i_1j_1}z_{j_1i_1})(z_{i_2j_2}z_{j_2i_2})\cdots(z_{i_mj_m}z_{j_mi_m})}{\cal P}_{\{i,j\}}(n,2m,m)\,,~~~~\label{poly}
\eea
where ${\cal P}_{\{i,j\}}(n,l,m)$ is defined in \eref{multi-trace-poly}, and we have arranged elements as
\bea
[X_k]_2=\left(
         \begin{array}{c|c}
         0& {1\over z_{i_kj_k}}\\
           \hline
           {1\over z_{j_ki_k}} & 0\\
         \end{array}
       \right)\,.
\eea
We want to emphasize that the multiple action of ${\cal T}_{ij}$ on ${\bf
Pf}'\Psi$ produce the structure ${\cal P}_{\{i,j\}}(n,l,m)$, which
is crucial for many theories. This is why it is called the trace
operator.

\subsection{Insertion operator}

The insertion operator is defined by \cite{Cheung:2017ems}
\bea
{\cal T}_{ikj}\equiv\partial_{k_i\epsilon_k}-\partial_{k_j\epsilon_k}\,.
\eea
As pointed out in  \cite{Cheung:2017ems}, ${\cal T}_{ikj}$ itself is
not a gauge invariant operator, but when it acts on objects obtained
after acting one trace operators, it is effectively gauge invariant.
Thus we  consider the effect of acting this operator on the
polynomial ${\sum_{\{i,j\}}}'{\cal P}_{\{i,j\}}(n,l,m)$ only.
According to discussions in \cite{Cheung:2017ems}, one should assume
that $k\in\{1,2,\cdots,n-l\}$ and $i,j\in{\rm Tr}_i$ to protect the
gauge invariance. For simplicity, we assume $i,j\in{\rm Tr}_m$ and
taking the expansion \eref{multi-trace-poly} where ${\rm Tr}_m$
has been deleted. This gauge choice will greatly simplify our
discussion, since with this choice $k_i\epsilon_k$ can appear in
\eref{multi-trace-poly} only through $C_{kk}$.

Initially, ${\bf Pf}[\Psi]_{n-l,i_1,j_1,\ldots,i_{m-1},j_{m-1}: n-l}$ is
\bea
{\bf Pf}[\Psi]_{n-l,i_1,j_1,\ldots,i_{m-1},j_{m-1}: n-l}=\sum_{\a\in\Pi}{\bf sgn}(\pi_{\a})[\Psi]_{a_1b_1}[\Psi]_{a_2b_2}\cdots[\Psi]_{a_{(n'+m')}b_{(n'+m')}}\,,~~~~\label{pfaffian-insertion}
\eea
where the definition in \eref{pfa} has been used. The element $[\Psi]_{a_ib_i}$ is at the $a_i^{\rm th}$ row and $b_i^{\rm th}$ column of the matrix $[\Psi]_{n-l,i_1,j_1,\ldots,i_{m-1},j_{m-1}: n-l}$, and we have defined $n'=n-l$, $m'=m-1$.
Since $\epsilon_k$ appears only in $C_{kk}$, when acting $\partial_{k_i\epsilon_k}$ on \eref{pfaffian-insertion}, only terms containing element $[\Psi]_{k,n'+2m'+k}$ (see the expression \eref{psi-aba}) can survive. Consider
such a term, the remaining part after the action corresponds to a partition of the
the  set $\{1,2,\cdots,2(n'+m')\}\setminus\{k,n'+2m'+k\}$, which has the length $2(n'+m'-1)$. Such a term appears in the  ${\bf Pf}[\Psi]_{n-l-1,i_1,j_1,\ldots,i_{m-1},j_{m-1}: n-l-1}$,
weighted by a different signature ${\bf sgn}(\pi_{\W\a})$, where  the
new matrix $[\Psi]_{n-l-1,i_1,j_1,\ldots,i_{m-1},j_{m-1}: n-l-1}$ is obtained from the original one $[\Psi]_{n-l,i_1,j_1,\ldots,i_{m-1},j_{m-1}: n-l}$ by deleting $k^{\rm th}$ and $(n'+2m'+k)^{\rm th}$ rows and columns,
and ${\bf sgn}(\pi_{\W\a})$ corresponds to the partition of the
length-$2(n'+m'-1)$ set. By comparing these two special partitions, where one belongs to the original
matrix and one belongs to the new one,
\bea
\a&=&\{(a_1,b_1),(a_2,b_2),\cdots,(k,n'+2m'+k),\cdots,(a_{(n'+m')},b_{(n'+m')})\},\nn
\W\a&=&\{(a_1,b_1),(a_2,b_2),\cdots,(a_{(n'+m'-1)},b_{(n'+m'-1)})\}\,,
\eea
one can get  ${\bf sgn}(\pi_{\a})={\bf sgn}(\pi_{\W\a})$ since $\W\a$ is obtained from $\a$ by deleting the pair $(k,n'+2m'+k)$.
Using above observation, when we sum all contributions together, we will have
\bea
\partial_{k_i\epsilon_k}{\bf Pf}[\Psi]_{n-l,i_1,j_1,\ldots,i_{m-1},j_{m-1}: n-l}={-1\over z_{ik}}{\bf Pf}[\Psi]_{n-l-1,i_1,j_1,\ldots,i_{m-1},j_{m-1}: n-l-1}\,.
\eea

Applying this result to \eref{multi-trace-poly}, we get immediately
\bea
{\cal T}_{ikj}\Big({\sum_{\{i,j\}}}'{\cal P}_{\{i,j\}}(n,l,m)\Big)&=&\Big({1\over z_{jk}}-{1\over z_{ik}}\Big)\Big({\sum_{\{i,j\}}}'{\cal P}_{\{i,j\}}(n,l+1,m)\Big)\nn
&=&{-z_{ij}\over z_{ik}z_{kj}}\Big({\sum_{\{i,j\}}}'{\cal P}_{\{i,j\}}(n,l+1,m)\Big)\,.~~~~\label{result-insertion}
\eea
Let us give a little bit explanation of  the result \eref{result-insertion}. There are two parts. The part
$\Big({\sum_{\{i,j\}}}'{\cal P}_{\{i,j\}}(n,l+1,m)\Big)$ means we have  added  a new element $k$ into the set ${\rm Tr}_m$. Here is no ordering of the set and every element is at the same footing. The ordering information comes
from the part ${-z_{ij}\over z_{ik}z_{kj}}$, especially the denominator factor $z_{ik}z_{kj}$ gives a line connecting
$i$ to $k$ and then $k$ to $j$, i.e., one has inserted the element $k$ between $i,j$.

To really achieve the goal, from Table \ref{tab:theories}, one can see that ${\sum_{\{i,j\}}}'{\cal P}_{\{i,j\}}(n,l,m)$
always appears together with a series of Parke-Taylor factors ${\cal C}_{{\rm Tr}_1}\cdots{\cal C}_{{\rm Tr}_m}$.
If the original ${\cal C}_{{\rm Tr}_m}$ contains $1/z_{ij}$, multiplying the factor $z_{ij}/(z_{ik}z_{jk})$ replaces it with $1/z_{ik}z_{kj}$, therefore implies the new color order $\{...ikj...\}$, i.e., the insertion of the element $k$ between $i,j$. This explanation tells us how to systematically insert elements into a trace one by one with a
well defined sequence of insertion operators.

Since the polynomial ${\sum_{\{i,j\}}}'{\cal P}_{\{i,j\}}(n,l,m)$ is independent of the choice
of the deleted rows and columns, assuming $i$ and $j$ belong to any other ${\rm Tr}_k$ will
lead to the same conclusion, although the calculation will be more complicate.

\subsection{Longitudinal operator}

The longitudinal operators are defined via \cite{Cheung:2017ems}
\bea
{\cal L}_i\equiv\sum_{j\neq i}k_i k_j\partial_{k_j\epsilon_i}\,,
\eea
and
\bea
{\cal L}_{ij}\equiv-k_i k_j\partial_{\epsilon_i\epsilon_j}\,.
\eea
Among these two, the ${\cal L}_{ij}$ is intrinsically gauge
invariant, but ${\cal L}_{i}$ is not\footnote{It is useful to
compare operators ${\cal L}_{ij}$ and ${\cal T}_{ij}$: they differ
by the factor $k_i\cdot k_j$, which turns the interaction into
derivatively coupling. Their common part, i.e.,
$\partial_{\epsilon_i\epsilon_j}$ plays the same role, i.e.,
"compactify".}. We now discuss the effects of acting them on the
reduced Pfaffian ${\bf Pf}'[\Psi]_{a,b:a}$.

We first consider the operator ${\cal L}_{ij}$. It turns $\epsilon_i\epsilon_j$ into
$k_ik_j$, and annihilates all other $\epsilon_iV$'s, $\epsilon_jV$'s. Using the observation that $\epsilon_i$
and $\epsilon_j$ can appear once and only once respectively, one can conclude that ${\cal L}_{ij}$ changes
the reduced Pfaffian of the matrix $[\Psi]_{a,b:a}$ as
\bea
{\cal L}_{ij}\,{\bf Pf}'\left(
         \begin{array}{c|c}
           ~~A_{(a+b)\times (a+b)}~~ &  C_{(a+b)\times a}  \\
           \hline
           -C^{\rm T}_{a\times (a+b)} & B_{a\times a}\\
         \end{array}
       \right)
       \Rightarrow
{\bf Pf}'\left(
         \begin{array}{c|c|c}
           ~~A_{(a+b)\times (a+b)}~~ &  C_{(a+b)\times (a-2)} & 0 \\
           \hline
           -C^{\rm T}_{(a-2)\times (a+b)} & B_{(a-2)\times(a-2)} & 0\\
           \hline
           0 & 0 & A_{2\times2}\\
         \end{array}
       \right)\,.~~~~\label{long1}
\eea
Next, we turn to the operator ${\cal L}_i$, which replaces every $k_j\epsilon_i$ with $k_jk_i$.
Under such replacement, the diagonal elements of the matrix $C$
become
\bea
C_{ii}\to-\sum_{l=1,\,l\neq i}^n{k_l \cdot k_i\over z_{li}}\,,
\eea
which will vanish due to the scattering equation. Thus, the effect of ${\cal L}_i$ is given by
\bea
{\cal L}_i\,{\bf Pf}'\left(
         \begin{array}{c|c}
           ~~A_{(a+b)\times (a+b)}~~ &  C_{(a+b)\times a}  \\
           \hline
           -C^{\rm T}_{a\times (a+b)} & B_{a\times a}\\
         \end{array}
       \right)
       \Rightarrow
{\bf Pf}'\left(
         \begin{array}{c|c|c}
           ~~A_{(a+b)\times (a+b)}~~ &  C_{(a+b)\times (a-2)} & A_{(a+b)\times2} \\
           \hline
           -C^{\rm T}_{(a-2)\times (a+b)} & B_{(a-2)\times(a-2)} & 0\\
           \hline
           A_{2\times (a+b)} & 0 & 0\\
         \end{array}
       \right)\,.~~~~\label{long2}
\eea

At this moment,  the meaning of \eref{long1} and \eref{long2} is not
clear. Actually, the longitudinal operators can not be performed
individually to generate any object belongs to physical integrands.
Instead, they should be used in a special manner, which will be
discussed  in the next section.

\section{Products of basic operators}
\label{secpro}

Using the products of basic operators, more operators will be
constructed. In this section, we will discuss these composed
operators, especially their action on the reduced Pfaffian ${\bf
Pf}'\Psi$, which is the fundamental building-block for the integrand
of gravity theory.

\subsection{Operator ${\cal T}[\a]$}

The operator ${\cal T}[\a]$ for a length-$m$ set
$\a=\{\a_1,\a_2,\cdots,\a_m\}$ is  defined as\footnote{We adopt the
convention in \cite{Cheung:2017ems} that the product of two
operators ${\cal O}_1\cdot{\cal O}_2$ acts on an amplitude as
$({\cal O}_1\cdot{\cal O}_2){\cal A} ={\cal O}_2{\cal O}_1{\cal A}$,
i.e., the operator ${\cal O}_1$ is performed at first, and ${\cal
O}_2$ secondly.} \cite{Cheung:2017ems}
\bea
{\cal T}[\a]\equiv{\cal T}_{\a_1\a_m}\cdot\prod_{i=2}^{m-1}{\cal T}_{\a_{i-1}\a_i\a_m}\,.
\eea
We now act this operator on ${\bf Pf}'\Psi$.
Firstly, performing ${\cal T}_{\a_1\a_m}$ gives
\bea
{\cal T}_{\a_1\a_m}\,{\bf Pf}'\Psi&=&{-1\over z_{\a_1\a_m}z_{\a_m\a_1}}z_{\a_1\a_m}{\bf Pf}'[\Psi]_{n-2,\a_1,\a_m:n-2}\nn
&=&{1\over z_{\a_1\a_m}z_{\a_m\a_1}}{\bf Pf}[\Psi]_{n-2:n-2}\,,
\eea
where \eref{result-trace-1} and $(-)^{(n-1)+n}=(-)$ have been used.
Then one can act ${\cal T}_{\a_1\a_2\a_m}$ on it, and use \eref{result-insertion} to get
\bea
{\cal T}_{\a_1\a_2\a_m}{\cal T}_{\a_1\a_m}\,{\bf Pf}'\Psi&=&{1\over z_{\a_1\a_m}z_{\a_m\a_1}}{-z_{\a_1\a_m}\over z_{\a_1\a_2}z_{\a_2\a_m}}{\bf Pf}[\Psi]_{n-3:n-3}\nn
&=&{-1\over z_{\a_1\a_2}z_{\a_2\a_m}z_{\a_m\a_1}}{\bf Pf}[\Psi]_{n-3:n-3}\,.
\eea
Similarly, one can obtain
\bea
{\cal T}_{\a_2\a_3\a_m}{\cal T}_{\a_1\a_2\a_m}{\cal T}_{\a_1\a_m}\,{\bf Pf}'\Psi&=&{-1\over z_{\a_1\a_2}z_{\a_2\a_m}z_{\a_m\a_1}}
{-z_{\a_2\a_m}\over z_{\a_2\a_3}z_{\a_3\a_m}}{\bf Pf}[\Psi]_{n-4:n-4}\nn
&=&{1\over z_{\a_1\a_2}z_{\a_2\a_3}z_{\a_3\a_m}z_{\a_m\a_1}}{\bf Pf}[\Psi]_{n-4:n-4}\,.
\eea
This procedure can be repeated recursively, and finally one will arrive
\bea {\cal T}[\a]\,{\bf Pf}'\Psi&=&{(-)^m\over
z_{\a_1\a_2}z_{\a_2\a_3}\cdots z_{\a_{m-1}\a_m}z_{\a_m\a_1}}{\bf
Pf}[\Psi]_{n-m:n-m}\nn &=&(-)^{m+1}{\cal C}_\a{\sum_{\{i,j\}}}'{\cal
P}_{\{i,j\}}(n,m,1)\,.~~~\label{single-trace} \eea
Above calculation is straightforward as long as $m\leq n-1$. The
case $m=n$ needs a careful treatment. When $m=n$, the final
insertion operator ${\cal T}_{\a_{n-2}\a_{n-1}\a_n}$ acts on the
Pfaffian of the $2\times2$ matrix $[\Psi]_{1:1}$ which is given as
\bea
[\Psi]_{1:1}=\left(
         \begin{array}{c|c}

           0 & C_{\a_{n-1},\a_{n-1}}\\
           \hline
           -C^{\rm T}_{\a_{n-1},\a_{n-1}} & 0\\
         \end{array}
       \right)\,.
\eea
The Pfaffian of this matrix is
\bea
{\bf Pf}[\Psi]_{1:1}&=&C_{\a_{n-1},\a_{n-1}}
=-\sum_{l=1,\,l\neq \a_{n-1}}^n{k_l\epsilon_{\a_{n-1}}\over z_{l,\a_{n-1}}}\,.
\eea
Applying ${\cal T}_{\a_{n-2}\a_{n-1}\a_n}$ on it, we get
\bea
{\cal T}[\a_1,\a_2,\cdots,\a_n]\,{\bf Pf}'\Psi={(-)^n\over z_{\a_1\a_2}z_{\a_2\a_3}\cdots z_{\a_{n-1}\a_n}z_{\a_n\a_1}}
=(-)^n{\cal C}_n\,.~~~~\label{rela-2}
\eea

The above result can be generalized to multi-trace cases ${\cal
T}[\a_1]\cdot{\cal T}[\a_2]\cdots$, via general relations
\eref{poly} and \eref{result-insertion}, with the constraint
$[\a_i]\cap[\a_j]=\emptyset$. Let us consider, for example,
\bea
{\cal T}[\a]\cdot{\cal T}[\b]&=&\Big({\cal T}_{\a_1\a_m}\cdot\prod_{i=2}^{m-1}{\cal T}_{\a_{i-1}\a_i\a_m}\Big)
\cdot\Big({\cal T}_{\b_1\b_l}\cdot\prod_{i=2}^{l-1}{\cal T}_{\b_{i-1}\b_i\b_l}\Big)\nn
&=&{\cal T}_{\a_1\a_m}\cdot{\cal T}_{\b_1\b_l}\cdot\Big(\prod_{i=2}^{m-1}{\cal T}_{\a_{i-1}\a_i\a_m}\Big)
\cdot\Big(\prod_{i=2}^{l-1}{\cal T}_{\b_{i-1}\b_i\b_l}\Big)\,.
\eea
The first step is using \eref{poly} to obtain
\bea
{\cal T}_{\b_1\b_l}{\cal T}_{\a_1\a_m}\,{\bf Pf}'\Psi
&=&\Big({-1\over z_{\b_1\b_l}z_{\b_l\b_1}}\Big)
\Big({-1\over z_{\a_1\a_m}z_{\a_m\a_1}}\Big){\sum_{\{i,j\}}}'{\cal P}_{\{i,j\}}(n,4,2)\,,
\eea
where
\bea
{\sum_{\{i,j\}}}'{\cal P}_{\{i,j\}}(n,4,2)={\cal P}_{\{i,j\}}(n,4,2)=z_{\b_1\b_l}{\bf Pf}[\Psi]_{n-4,\b_1,\b_l:n-4}
=z_{\a_1\a_m}{\bf Pf}[\Psi]_{n-4,\a_1,\a_m:n-4}\,.
\eea
Secondly, one can use \eref{result-insertion} to get
\bea
& &{\cal T}_{\a_{m-2}\a_{m-1}\a_m}\cdots{\cal T}_{\a_{2}\a_{3}\a_m}{\cal T}_{\a_{1}\a_{2}\a_m}\Big({-1\over z_{\a_1\a_m}z_{\a_m\a_1}}\Big){\sum_{\{i,j\}}}'{\cal P}_{\{i,j\}}(n,4,2)\nn
&=&{\cal T}_{\a_{m-2}\a_{m-1}\a_m}\cdots{\cal T}_{\a_{2}\a_{3}\a_m}\Big({1\over z_{\a_1\a_2}z_{\a_2\a_m}z_{\a_m\a_1}}\Big){\sum_{\{i,j\}}}'{\cal P}_{\{i,j\}}(n,5,2)\nn
& &~~~~~~~~~~~~~\cdots\nn
&=&(-)^{m+1}{\cal C}_{\a}{\sum_{\{i,j\}}}'{\cal P}_{\{i,j\}}(n,2+m,2)\,.
\eea
Thirdly, we use \eref{result-insertion} again to obtain
\bea
& &{\cal T}_{\b_{l-2}\a_{l-1}\a_l}\cdots{\cal T}_{\b_{2}\b_{3}\b_l}{\cal T}_{\b_{1}\b_{2}\b_l}\Big({-1\over z_{\b_1\b_l}z_{\b_l\b_1}}\Big){\sum_{\{i,j\}}}'{\cal P}_{\{i,j\}}(n,2+m,2)\nn
&=&(-)^{l+1}{\cal C}_{\b}{\sum_{\{i,j\}}}'{\cal P}_{\{i,j\}}(n,l+m,2)\,.
\eea
Combining them together we get
\bea
{\cal T}[\a]\cdot{\cal T}[\b]\,{\bf Pf}'\Psi=(-)^{m+l+2}{\cal C}_{\a}{\cal C}_{\b}{\sum_{\{i,j\}}}'{\cal P}_{\{i,j\}}(n,l+m,2)\,.
\eea
Now one can see the recursive pattern that
\bea
{\cal T}[\a_1]\cdot{\cal T}[\a_2]\cdots{\cal T}[\a_k]\,{\bf Pf}'\Psi=(-)^{k+\sum |\a_i|}\Big(\prod{\cal C}_{\a_i}\Big){\sum_{\{i,j\}}}'{\cal P}_{\{i,j\}}(n,\sum |\a_i|,k)\,,~~~~\label{rela-3}
\eea
where $|\a_k|$ denotes the length of the set $\a_k$.

\subsection{Operator ${\cal L}\cdot{\cal T}_{ab}$}

The operator ${\cal L}$ is defined through longitudinal operators as \cite{Cheung:2017ems}
\bea {\cal L}\equiv\prod_i{\cal L}_i={\cal \W L}+\cdots\,,~~~~{\rm with}~{\cal \W L}\equiv
\sum_{\rho\in {\rm pair}} \prod_{i,j\in\rho}{\cal
L}_{ij}.~~~\label{LT} \eea
The expression \eref{LT} means that  at the algebraic level, the
effect of $\prod_i{\cal L}_i$ is different from that of
$\sum_{\rho\in {\rm pair}}\prod_{i,j\in\rho}{\cal L}_{ij}$.
However, if one consider the combination ${\cal L}\cdot{\cal
T}_{ab}\,{\bf Pf}'\Psi$, and let subscripts of ${\cal L}_i$'s and ${\cal L}_{ij}$'s
run through all nodes in $\{1,2,\cdots,n\}\setminus\{a,b\}$, the
effects of $\prod_i{\cal L}_i$ and $\sum_{\rho\in {\rm
pair}}\prod_{i,j\in\rho}{\cal L}_{ij}$ are same, give a result which
has a meaningful explanation.

Let us first study the effect of the operation ${\cal \W
L}\cdot{\cal T}_{ab}\,{\bf Pf}'\Psi$. Since ${\cal \W L}$ and
${\cal T}_{ab}$ are commutable, i.e., ${\cal\W L}\cdot{\cal
T}_{ab}={\cal T}_{ab}\cdot{\cal \W L}$, we will apply the operator
${\cal T}_{ab}$ on ${\bf Pf}'\Psi$ firstly to get
\eref{result-trace-1}, then act ${\cal \W L}$ on it. It is straightforward to see $\sum_{\rho\in {\rm
pair}}\prod_{i,j\in\rho}{\cal L}_{ij}$ changes the matrix
\eref{result-M} into
\bea
\Psi' = \left(
         \begin{array}{c|c|c}
           ~~A_{n\times n}~~ &  0 & 0 \\
           \hline
          0 & -A_{(n-2)\times(n-2)} & 0\\
           \hline
           0 & 0 & X_{2\times2}\\
         \end{array}
       \right)\,,
\eea
due to the previous result \eref{long1}.  The Pfaffian of the matrix
\bea
\left(
         \begin{array}{c|c}

           -A_{(n-2)\times(n-2)} & 0\\
           \hline
           0 & X_{2\times2}\\
         \end{array}
       \right)\,,
\eea
is just $(-)^{a+b}{\bf Pf}'(-A)=(-)^{{n\over2}-1+a+b}{\bf Pf}'A$, thus
\bea
\sum_{\rho\in {\rm pair}}\prod_{i,j\in\rho}{\cal L}_{ij}\cdot{\cal T}_{ab}\,{\bf Pf}'\Psi
={\bf Pf}'\Psi'=(-)^{{n\over2}-1+a+b}\Big({\bf Pf}'A\Big)^2\,.
\eea

Next we consider the effect of acting $\prod_i{\cal L}_i$
on ${\cal T}_{ab}\,{\bf Pf}'\Psi$.
Using \eref{long2} we know the operator $\prod_i{\cal L}_i$ turns
the matrix \eref{result-M} into
\bea
\Psi''= \left(
         \begin{array}{c|c|c}
           ~~A_{n\times n}~~ &  A_{n\times(n-2)} & 0 \\
           \hline
          A_{(n-2)\times n} & 0 & 0\\
           \hline
           0 & 0 & X_{2\times2}\\
         \end{array}
       \right)\,,
\eea
thus the reduced Pfaffian is
\bea {\bf Pf}'\Psi''= {\bf Pf}'\W A\,{\bf Pf}[X]_2\,,\eea
where
\bea
\W A\equiv \left(
         \begin{array}{c|c}
           ~~A_{n\times n}~~ &  A_{n\times(n-2)} \\
           \hline
          A_{(n-2)\times n} & 0 \\
         \end{array}
       \right)\,.
\eea
To compute the reduced Pfaffian of $\W A$, we choose $a^{\rm th}$
and $b^{\rm th}$ rows  and columns of $A_{n\times n}$ to be removed. Furthermore one
can use the relation that for the matrix $S$ with the block
structure
\bea
S=\left(
         \begin{array}{c|c}

           M & Q\\
           \hline
           -Q^{\rm T} & N\\
         \end{array}
       \right)\,,
\eea
when $M$ is invertible, the Pfaffian of $S$ satisfies
\bea
{\bf Pf}S={\bf Pf}M\,{\bf Pf}(N+Q^{\rm T}M^{-1}Q)\,.
\eea
Using this, the reduced Pfaffian of $\W A$
 can be calculated as
\bea
{\bf Pf}'\W A&= &{(-)^{a+b}\over z_{ab}}{\bf Pf}A_{(n-2)\times(n-2)}\,{\bf Pf}\Big(0+A_{(n-2)\times(n-2)}^{\rm T}A_{(n-2)\times(n-2)}^{-1}A_{(n-2)\times(n-2)}\Big)\nn
&=&{(-)^{a+b}\over z_{ab}}{\bf Pf}A_{(n-2)\times(n-2)}\,{\bf Pf}\Big(0+A_{(n-2)\times(n-2)}^{\rm T}\Big)\nn
&=&{(-)^{a+b}\over z_{ab}}{\bf Pf}A_{(n-2)\times(n-2)}\,{\bf Pf}\Big(-A_{(n-2)\times(n-2)}\Big)\nn
&=&(-)^{{n\over2}-1+a+b}z_{ab}\Big({\bf Pf}'A\Big)^2\,,
\eea
Putting it back we obtain
\bea
\prod_i{\cal L}_i\cdot{\cal T}_{ab}\,{\bf Pf}'\Psi
={\bf Pf}'\Psi''=(-)^{{n\over2}-1+a+b}\Big({\bf Pf}'A\Big)^2\,.
\eea
Above calculations show that
\bea {\cal L}\cdot{\cal T}_{ab}\,{\bf Pf}'\Psi={\cal \W L}\cdot{\cal
T}_{ab}\,{\bf Pf}'\Psi =(-)^{{n\over2}-1+a+b}\Big({\bf
Pf}'A\Big)^2\,.~~~~\label{rela-4} \eea
It is worth to notice that this result is independent of the choice of $a$ and $b$.

\subsection{New operators ${\cal T}_{X_{2m}}$ and ${\cal T}_{{\cal X}_{2m}}$ }

As can be seen in  Table \ref{tab:theories}, the CHY-integrands for
several  theories require the ingredients ${\bf Pf}[X]_{2m}$ and ${\bf Pf}[{\cal X}]_{2m}$. These objects can also be created from
the original matrix $\Psi$ via appropriate operators. Now we give the definition of these new operators.

For a given length-$2m$ set $I$, we define a new operator as
\bea
{\cal T}_{X_{2m}}\equiv\sum_{\rho\in {\rm pair}}\prod_{i,j\in\rho}{\cal T}_{i_kj_k}\,.
~~~\label{4.28}\eea
Here the set of pairs $\{(i_1,j_1),(i_2,j_2),\cdots,(i_m,j_m)\}$ is
a partition of $I$ with conditions $i_1<i_2<...<i_m$ and
$i_t<j_t,~\forall t$. Using the result in \eref{poly} as well as the
\eref{pfa}, one can conclude that the operator ${\cal T}_{X_{2m}}$
generates a new matrix
\bea
\W\Psi^\ast = \left(
         \begin{array}{c|c|c}
           ~~A_{n\times n}~~ &  -C^{\rm T}_{n\times(n-2m)} & 0 \\
           \hline
           C_{(n-2m)\times n} & B_{(n-2m)\times(n2-m)} & 0\\
           \hline
           0 & 0 & X_{2m\times2m}\\
         \end{array}
       \right)\,,~~~~~\label{result-M'}
\eea
such that acting ${\cal T}_{X_{2m}}$ on the reduced Pfaffian ${\bf
Pf}'\Psi$ gives
\bea
{\cal T}_{X_{2m}}\,{\bf Pf}'\Psi={\bf Pf}'\W\Psi^\ast={\bf Pf}'[\Psi]_{n-2m,2m:n-2m}{\bf Pf}[X]_{2m}\,,~~~~\label{rela-5}
\eea
which provides the desired building block ${\bf Pf}[X]_{2m}$.

By similar argument, we can also define the operator ${\cal
T}_{{\cal X}_{2m}}$ as
\bea
{\cal T}_{{\cal X}_{2m}}\equiv\sum_{\rho\in {\rm pair}}\prod_{i,j\in\rho}\delta^{I_{i_k},I_{j_k}}{\cal T}_{i_kj_k}\,,
~~~\label{4.31}\eea
which is the generalization of ${\cal T}_{{\cal X}_{2m}}$. The $\delta^{I_{i_k},I_{j_k}}$'s turn the matrix $[X]_{2m}$ into $[{\cal X}]_{2m}$,
therefore we get
\bea
{\cal T}_{{\cal X}_{2m}}\,{\bf Pf}'\Psi={\bf Pf}'[\Psi]_{n-2m,2m:n-2m}{\bf Pf}[{\cal X}]_{2m}\,,~~~~\label{rela-6}
\eea
which gives the required building block ${\bf Pf}[{\cal X}]_{2m}$. Before ending this part, we want to emphasize
one important point:  since ${\cal T}_{ij}$
is intrinsically gauge invariant, so are ${\cal T}_{{ X}_{2m}}$ and ${\cal T}_{{\cal X}_{2m}}$.

\section{Unifying relations for amplitudes}
\label{secunify}

With preparations in previous sections, we are ready to exhibit
relations between amplitudes. As discussed in \S\ref{secintro}, the
idea is, differential operators are commutable with the integration
over complex variables $z_i$'s, thus the effects of acting them on
amplitudes can be realized as acting on corresponding CHY-integrands, and vice versa. Our previous calculations have explicitly established the
relation between two approaches in \cite{Cachazo:2014xea} and
\cite{Cheung:2017ems}. In this section, we will apply our results
 in sections \S\ref{secbasic} and \S\ref{secpro} to write down
relations between different scattering amplitudes, as did in
\cite{Cachazo:2014xea} and \cite{Cheung:2017ems}.

\subsection{The unified web}

Now we act the operators on CHY integrands for various theories to
get the unifying relations for amplitudes. The starting point is the
formulation for the  gravity theory. The reason is,
all operators decrease the spins of external particles, thus the
unified web must start from the amplitudes for gravitons which carry
highest spins. The integrand of  gravity theory is shown in the
first line of Table \ref{tab:theories}, two parts ${\cal I}_L$ and
${\cal I}_R$ depend on two independent sets of polarization vectors
$\{\epsilon\}$ and $\{\W\epsilon\}$, respectively. Since all
operators are defined through partial differentials of some Lorentz
invariants contain polarization vectors, it is natural to restrict
the effect of them on the ${\cal I}_L$ part (or equivalently the
${\cal I}_R$ part), by defining operators via $\epsilon$ (or
$\W\epsilon$).  Performing operators on the ${\cal I}_L$ part and
using \eref{poly}, \eref{rela-2}, \eref{rela-3}, \eref{rela-4},
\eref{rela-5} and \eref{rela-6}, after comparing with the middle column of Table
\ref{tab:theories}, we get following relations:
\bea
{\cal A}^{{\rm EYM}}&=&{\cal T}[{\rm Tr}_1]\cdots{\cal T}[{\rm Tr}_m]\,{\cal A}^{{\rm G}}\,,\nn
{\cal A}^{{\rm YM}}&=&{\cal T}[i_1\cdots i_n]\,{\cal A}^{{\rm G}}\,,\nn
{\cal A}^{{\rm EM}}&=&{\cal T}_{X_{2m}}\,{\cal A}^{{\rm G}}\,,\nn
{\cal A}^{{\rm EM}}_{{\rm flavor}}&=&{\cal T}_{{\cal X}_{2m}}\,{\cal A}^{{\rm G}}\,,\nn
{\cal A}^{{\rm BI}}&=&{\cal L}\cdot{\cal T}[ab]\,{\cal A}^{{\rm G}}\,,~~\label{A-rela-1}
\eea
up to an overall sign. Here ${\cal A}^{{\rm G}}$, ${\cal A}^{{\rm EYM}}$, ${\cal A}^{{\rm YM}}$, ${\cal A}^{{\rm EM}}$,
${\cal A}^{{\rm EM}}_{{\rm flavor}}$, ${\cal A}^{{\rm BI}}$ denote amplitudes of gravity theory,
Einstein-Yang-Mills theory, pure Yang-Mills theory, Einstein-Maxwell theory, Einstein-Maxwell theory that photons carry flavors,
Born-Infeld theory, respectively.

For the pure Yang-Mills integrand, there is only one copy ${\bf pf}'\,\Psi$ depends on polarization
vectors, thus operators can be performed directly. Starting from the pure Yang-Mills integrand,
we obtain relations:
\bea
{\cal A}^{{\rm YMS}}&=&{\cal T}[{\rm Tr}_1]\cdots{\cal T}[{\rm Tr}_m]\,{\cal A}^{{\rm YM}}\,,\nn
{\cal A}^{{\rm YMS}}_{{\rm special}}&=&{\cal T}_{{\cal X}_{2m}}\,{\cal A}^{{\rm YM}}\,,\nn
{\cal A}^{{\rm BS}}&=&{\cal T}[i_1\cdots i_n]\,{\cal A}^{{\rm YM}}\,,\nn
{\cal A}^{{\rm NLSM}}&=&{\cal L}\cdot{\cal T}[ab]\,{\cal A}^{{\rm YM}}\,,\nn
{\cal A}^{\phi^4}&=&{\cal T}_{X_{n}}\,{\cal A}^{{\rm YM}}\,,~~\label{A-rela-2}
\eea
up to an overall sign, where ${\cal A}^{{\rm YMS}}$, ${\cal A}^{{\rm YMS}}_{{\rm special}}$, ${\cal A}^{{\rm BS}}$, ${\cal A}^{{\rm NLSM}}$,
${\cal A}^{\phi^4}$ denote amplitudes of Yang-Mills-scalar theory, special Yang-Mills-scalar theory, bi-adjoint scalar theory, non-linear
sigma model, as well as $\phi^4$ theory, respectively. Notice that the amplitude of $\phi^4$ theory is generated
via a special ${\cal T}_{X_{2m}}$ that $2m=n$.

Applying operators on the Born-Infeld integrand, we get relations:
\bea
{\cal A}^{{\rm DBI}}_{{\rm ex}}&=&{\cal T}[{\rm Tr}_1]\cdots{\cal T}[{\rm Tr}_m]\,{\cal A}^{{\rm BI}}\,,\nn
{\cal A}^{{\rm DBI}}&=&{\cal T}_{{\cal X}_{2m}}\,{\cal A}^{{\rm BI}}\,,\nn
{\cal A}^{{\rm NLSM}}&=&{\cal T}[i_1\cdots i_n]\,{\cal A}^{{\rm BI}}\,,\nn
{\cal A}^{{\rm SG}}&=&{\cal L}\cdot{\cal T}[ab]\,{\cal A}^{{\rm BI}}\,,~~\label{A-rela-3}
\eea
up to an overall sign, where ${\cal A}^{{\rm DBI}}_{{\rm ex}}$, ${\cal A}^{{\rm DBI}}$, ${\cal A}^{{\rm NLSM}}$, ${\cal A}^{{\rm SG}}$
denote amplitudes of extended Dirac-Born-Infeld theory, Dirac-Born-Infeld theory, non-linear sigma model,
special Galileon theory, respectively.

Our results \eref{A-rela-1}, \eref{A-rela-2} and \eref{A-rela-3}, gives not only unified relations presented in
\cite{Cheung:2017ems}, but also other relations among theories having  CHY representations in
\cite{Cachazo:2014xea}. We want to remark that a result in this paper is different from the one
in \cite{Cheung:2017ems}, i.e., the Einstein-Maxwell theory: their differential operator
is just one term of the operator  ${\cal T}_{X_{2m}}$ defined in \eref{4.28}.

Relations presented above can be organized into Table \ref{tab:unifying}.
\begin{table}[!h]
    \begin{center}
        \begin{tabular}{c|c}
            Amplitude& Operator acts on ${\cal A}^{{\rm G}}(\epsilon,\W\epsilon,k)$ \\
            \hline
            ${\cal A}^{{\rm EYM}}(\epsilon,\W\epsilon,k)$ & ${\cal T}^{\epsilon}[{\rm Tr}_1]\cdots{\cal T}^{\epsilon}[{\rm Tr}_m]$ \\
            ${\cal A}^{{\rm YM}}(\W\epsilon,k)$&  ${\cal T}^{\epsilon}[i_1\cdots i_n]$ \\
            ${\cal A}^{{\rm EM}}(\epsilon,\W\epsilon,k)$ & ${\cal T}^{\epsilon}_{X_{2m}}$\\
            ${\cal A}^{{\rm EM}}_{{\rm flavor}}(\epsilon,\W\epsilon,k)$&  ${\cal T}^{\epsilon}_{{\cal X}_{2m}}$ \\
            ${\cal A}^{{\rm BI}}(\W\epsilon,k)$ & ${\cal L}^{\epsilon}\cdot{\cal T}^{\epsilon}[ab]$ \\
            ${\cal A}^{{\rm YMS}}(\W\epsilon,k)$ & ${\cal T}^{\epsilon}[i_1\cdots i_n]\cdot\Big({\cal T}^{\W\epsilon}[{\rm Tr}_1]\cdots{\cal T}^{\W\epsilon}[{\rm Tr}_m]\Big)$ \\
            ${\cal A}^{{\rm YMS}}_{{\rm special}}(\W\epsilon,k)$ &  ${\cal T}^{\epsilon}[i_1\cdots i_n]\cdot{\cal T}^{\W\epsilon}_{{\cal X}_{2m}}$ \\
            ${\cal A}^{{\rm BS}}(k)$ &  ${\cal T}^{\epsilon}[i_1\cdots i_n]\cdot{\cal T}^{\W\epsilon}[i_1'\cdots i_n']$ \\
            ${\cal A}^{{\rm NLSM}}(k)$ & ${\cal T}^{\epsilon}[i_1\cdots i_n]\cdot\Big({\cal L}^{\W\epsilon}\cdot{\cal T}^{\W\epsilon}_{a'b'}\Big)$ \\
            ${\cal A}^{\phi^4}(k)$ &  ${\cal T}^{\epsilon}[i_1\cdots i_n]\cdot {\cal T}^{\W\epsilon}_{X_n}$ \\
            ${\cal A}^{{\rm DBI}}_{{\rm ex}}(\W\epsilon,k)$ &  $\Big({\cal L}^{\epsilon}\cdot{\cal T}^{\epsilon}[ab]\Big)\cdot\Big({\cal T}^{\W\epsilon}[{\rm Tr}_1]\cdots{\cal T}^{\W\epsilon}[{\rm Tr}_m]\Big)$ \\
            ${\cal A}^{{\rm DBI}}(\W\epsilon,k)$ &$\Big({\cal L}^{\epsilon}\cdot{\cal T}^{\epsilon}[ab]\Big)\cdot{\cal T}^{\W\epsilon}_{{\cal X}_{2m}}$ \\
            ${\cal A}^{{\rm SG}}(k)$ &  $\Big({\cal L}^{\epsilon}\cdot{\cal T}^{\epsilon}[ab]\Big)\cdot\Big({\cal L}^{\W\epsilon}\cdot{\cal T}^{\W\epsilon}[a'b']\Big)$ \\
        \end{tabular}
    \end{center}
    \caption{\label{tab:unifying}Unifying relations}
\end{table}
In this table the notations ${\cal T}^{\epsilon}[{\rm Tr}_i]$ and ${\cal T}^{\W\epsilon}[{\rm Tr}_i]$ means
two operators are defined through two independent sets of polarization vectors $\{\epsilon\}$ and $\{\W\epsilon\}$
respectively, and so do notations of other operators. If one add the identical operator $\mathbb{I}$ into
the set of operators, Table \ref{tab:unifying} can be summarized as
\bea
{\cal A}^{{\rm other}}={\cal O}^{\epsilon}\cdot{\cal O}^{\W\epsilon}\,{\cal A}^{{\rm G}}(\epsilon,\W\epsilon,k)\,,~~~~\label{LR}
\eea
where ${\cal O}^{\epsilon}$ and ${\cal O}^{\W\epsilon}$ denote operators which are defined through $\{\epsilon\}$ and $\{\W\epsilon\}$
respectively. Since the manifest double copy structure of the CHY integrands, ${\cal O}^{\epsilon}$ and ${\cal O}^{\W\epsilon}$ are applied on two copies independently at the integrand-level.

\subsection{Other relations}

Differential operators connect not only amplitudes from different theories, but also amplitudes of
same type of theory.
For example, let us consider the Einstein-Yang-Mills theory. Let us start from a $(m+n)$-point color-ordered amplitude
${\cal A}^{\rm EYM}(i_1^h,\cdots, i_m^h;j_1^g,\cdots, j_n^g)$,
where $h$ and $g$ denote gravitons and gluons respectively with the color
order of gluons  as $\{j_1,j_2,\cdots, j_n\}$.  Using the relation \eref{result-insertion},
one can act insertion operators
to turn gravitons into gluons at any desired positions, such as  following:
\bea
{\cal A}^{\rm EYM}(i_3^h,\cdots, i_m^h;j_1^g,i_1^g,i_2^g,j_2^g,\cdots, j_n^g)
&=&{\cal T}_{i_1i_2j_2}\,{\cal T}_{j_1i_1j_2}\,{\cal A}^{\rm EYM}(i_1^h,\cdots, i_m^h;j_1^g,\cdots, j_n^g)\,,\nn
{\cal A}^{\rm EYM}(i_3^h,\cdots, i_m^h;j_1^g,i_1^g,j_2^g,i_2^g,j_3^g,\cdots, j_n^g)
&=&{\cal T}_{j_2i_2j_3}\,{\cal T}_{j_1i_1j_2}\,{\cal A}^{\rm EYM}(i_1^h,\cdots, i_m^h;j_1^g,\cdots, j_n^g)\,,
\eea
In above expressions, we have turned two gravitons into gluons, with different orderings:
the first one with ordering  $\{j_1,i_1,i_2,j_2,\cdots, j_n\}$ and the second one, $\{j_1,i_1,j_2,i_2,j_3,\cdots, j_n\}$,
respectively. Situations for other theories can be analyzed similarly.

One can also seek amplitudes for other theories beyond these given in Table \ref{tab:unifying},
by acting on the amplitude of gravity theory via other combinations of differential operators.
The operator ${\cal O}^{\epsilon}$ in \eref{LR} has $6$ choices which are
$\mathbb{I}$, ${\cal T}[{\rm Tr}_1]\cdots{\cal T}[{\rm Tr}_m]$, ${\cal T}[i_1\cdots i_n]$, ${\cal T}_{X_{2m}}$,
${\cal T}_{{\cal X}_{2m}}$, ${\cal L}\cdot{\cal T}_{ab}$, and so does ${\cal O}^{\W\epsilon}$.
Thus, starting from the CHY-integrand of gravity theory, there are $21$ kinds of CHY-integrands can be obtained by performing operators.
We now list the remaining $8$ cases as following:
\bea
& &{\cal T}^\epsilon_{X_{2m}}\cdot{\cal T}^{\W\epsilon}_{X'_{2m'}}\,,~~~~~~
{\cal T}^{\epsilon}[i_1\cdots i_n]\cdot{\cal T}^{\W\epsilon}_{X_{2m}}\,,~~~~~~
\Big({\cal L}^{\epsilon}\cdot{\cal T}^{\epsilon}[ab]\Big)\cdot{\cal T}^{\W\epsilon}_{X_{2m}}\,,~~~~~~
\Big({\cal T}^{\epsilon}[{\rm Tr}_1]\cdots{\cal T}^{\epsilon}[{\rm Tr}_m]\Big)\cdot{\cal T}^{\W\epsilon}_{X_{2m}}\,,\nn
& &\Big({\cal T}^{\epsilon}[{\rm Tr}_1]\cdots{\cal T}^{\epsilon}[{\rm Tr}_m]\Big)\cdot {\cal T}^{\W\epsilon}_{{\cal X}_{2m}}\,,~~~~
\Big({\cal T}^{\epsilon}[{\rm Tr}_1]\cdots{\cal T}^{\epsilon}[{\rm Tr}_m]\Big) \cdot \Big({\cal T}^{\W\epsilon}[{\rm Tr}_{1'}]\cdots{\cal T}^{\W\epsilon}[{\rm Tr}_{m'}]\Big)\,,\nn
& & {\cal T}^{\epsilon}_{{\cal X}_{2m}}\cdot {\cal T}^{\W\epsilon}_{X_{2m}}\,,~~~~
{\cal T}^{\epsilon}_{{\cal X}_{2m}}\cdot {\cal T}^{\W\epsilon}_{{\cal X}_{2m}}\,.~~~~\label{new-rela}
\eea
Using results in \S\ref{secpro}, one can get the corresponding integrands generated by them. If some of these integrands correspond to physical amplitudes, then new unifying relations occurs.
The complete analyis of various combinations in \eref{new-rela} is beyond the scope of this note and we will leave it to future work. Here we just give some brief discussions.

For the first case ${\cal T}^{\epsilon}_{X_{2m}}\cdot {\cal T}^{\W\epsilon}_{X'_{2m'}}$, when
$2m=2m'=n$, it yields the integrand
\bea
{\cal T}^{\epsilon}_{X_{n}}\cdot {\cal T}^{\W\epsilon}_{X_{n}}\,{\cal I}^{{\rm G}}(\epsilon,\W\epsilon,k,z)
=\Big({\bf Pf}'A_n\,{\bf Pf}[X]_n\Big)\Big({\bf Pf}'A_n\,{\bf Pf}[X]_n\Big)\,,~~~~\label{EMS}
\eea
where ${\cal I}^{{\rm G}}(\epsilon,\W\epsilon,k,z)$ denotes the integrand for gravity theory.
This is the integrand for Einstein-Maxwell-scalar theory, with all external particles are scalars \cite{Cachazo:2016njl}. This result is just a special case with $n=2m=2m'$. As we have emphasized, since
${\cal T}^{\epsilon}_{X_{n}}$ is intrinsically gauge invariant, we can take any length for this operator.
Furthermore, the role of ${\cal T}_{ij}$ is just to do the dimension reduction. With this understanding, one can see that
for general $m$ and $m'$ the  ${\cal T}^{\epsilon}_{X_{2m}}\cdot {\cal T}^{\W\epsilon}_{X'_{2m'}}\,{\cal A}^{{\rm G}}(\epsilon,\W\epsilon,k)$ will give the theory obtained from gravity theory by dimension reduction, i.e., the general Einstein-Maxwell-scalar amplitudes, whose external particles can be either
gravitons, photons, as well as scalars, i.e.,
\bea
{\cal A}^{{\rm EMS}}(\epsilon,\W\epsilon,k)={\cal T}^{\epsilon}_{X_{2m}}\cdot {\cal T}^{\W\epsilon}_{X'_{2m'}}\,{\cal A}^{{\rm G}}(\epsilon,\W\epsilon,k)\,.
\eea
For the second case ${\cal T}^{\epsilon}[i_1\cdots i_n]\cdot {\cal T}^{\W\epsilon}_{X_{2m}}$,
when $2m=n$, we get
\bea
{\cal T}^{\epsilon}[i_1\cdots i_n]\cdot {\cal T}^{\W\epsilon}_{X_{n}}\,{\cal I}^{{\rm G}}(\epsilon,\W\epsilon,k,z)
={\cal C}_n{\bf Pf}'A\,{\bf Pf}[X]_{n}\,,
\eea
which is the $\phi^4$ theory. Again, the operator ${\cal T}^{\W\epsilon}_{X_{2m}}$ can be any length. When $2m<n$,
we get the theory obtained by doing dimension reduction from Yang-Mills theory, which is the special Yang-Mills-Scalar theory
\bea
{\cal T}^{\epsilon}[i_1\cdots i_n]\cdot {\cal T}^{\W\epsilon}_{X_{2m}}\,{\cal I}^{{\rm G}}(\epsilon,\W\epsilon,k,z)
={\cal C}_n{\bf Pf}'[\Psi]_{n-2m,2m:n-2m}{\bf Pf}[X]_{2m}\,.~~~~\label{YM-phi4}
\eea
Here if we replace ${\cal T}^{\W\epsilon}_{X_{2m}}$ by $ {\cal T}^{\W\epsilon}_{{\cal X}_{2m}}$, we will
get the special Yang-Mills-Scalar theory with multiple kinds of scalars, as can be seen in Table \ref{tab:unifying}.

Other cases in \eref{new-rela} can be discussed similarly. One can obtain more  possible integrands via products
${\cal O}^\epsilon={\cal O}^\epsilon_1\cdots {\cal O}^\epsilon_a$ and ${\cal O}^{\W\epsilon}={\cal O}^{\W\epsilon}_1\cdots {\cal O}^{\W\epsilon}_b$. In general, for any ${\cal O}^{\epsilon}\cdot {\cal O}^{\W\epsilon}\,{\cal I}^{{\rm G}}(\epsilon,\W\epsilon,k,z)$, information of external particles such as spins and gauge structures can be read out directly from the obtained integrand, but pin down  the form of interaction
is a hard work.

\section{Summary and discussion}
\label{secconclu}

To summarize, we have provided manifest connection between two approaches, i.e., the differential operator  in \cite{Cheung:2017ems} and various manipulations (such as compactification and squeezing procedures)
in \cite{Cachazo:2014xea}. Using this connection, by acting differential operators on the CHY integrand of  gravity theory, one can
systematically derive  unifying relations for amplitudes of various theories, include Einstein gravity, Einstein-Yang-Mills theory, Einstein-Maxwell theory, pure Yang-Mills theory, Yang-Mills-scalar theory, Born-Infeld theory,
Dirac-Born-Infeld theory and its extension, bi-adjoint scalar theory, $\phi^4$ theory, non-linear sigma model,
as well as special Galileon theory. Along the line,  all unifying relations
in \cite{Cheung:2017ems} have been reproduced, and all theories which have CHY representations in \cite{Cachazo:2014xea}
have been included in the unified web. We have also discussed other new relations for amplitudes, which are indicated by our method.

The manifest double copy structure of the CHY integrand permits two sets of operators ${\cal O}^\epsilon$ and ${\cal O}^{\W\epsilon}$
to be applied independently. This advantage simplifies the derivation: it is sufficient to consider the effects of
acting operators on the reduced Pfaffian ${\bf Pf}'\Psi$.

A natural question will be, why these operators? From discussions in \cite{Cheung:2017ems}, one critical condition is the gauge symmetry.
The trace operators protect the gauge invariance while others do not. This is why the insertion and longitudinal operators should
be performed after the trace operator. There are other operators, such as ${\cal T}_{ijkl}$, have not been used
in the construction. Thus it will be interesting to consider broader form of differential operators. Furthermore, how
to understand these physical conditions from the point of view of CHY formulae is also important.

Our result can also be used to other studies. For example,
recent studies \cite{Fu:2017uzt,Teng:2017tbo,Du:2017gnh} have shown how to
expand the Einstein-Yang-Mills amplitudes by the Yang-Mills ones. If one act the differential operator at both sides
of the expansion, a differential equation connecting amplitudes of two different theories will be obtained. Solving this
differential equation (or doing the integration), we should find amplitudes for particles with higher spins from
other ones with lower spins. This is opposite to current construction of united web by starting from highest spin state, i.e., gravitons.

\section*{Acknowledgments}

We would thank Ellis Ye Yuan, Rijun Huang and Fei Teng for useful discussions. We also thank Ellis Ye Yuan
for comment on the draft. This
work is supported by Qiu-Shi Funding and Chinese NSF funding under
contracts No.11575156 and No.11805163.


\end{document}